\documentclass{ws-procs10x7}

\begin{document}

\title{Search for Higgs and New Phenomena at Colliders}

\author{Stephan Lammel}

\address{Fermi National Accelerator Laboratory, Batavia, Illinois 60510, USA\\
         E-mail: lammel@fnal.gov}

\twocolumn[\maketitle\abstract{
The present status of searches for the Higgs boson(s) and new phenomena
is reviewed. The focus is on analyses and results from the current runs
of the HERA and Tevatron experiments. The LEP experiments have released
their final combined MSSM Higgs results for this conference. Also included
are results from sensitivity studies of the LHC experiments and lepton
flavour violating searches from the B factories, KEKB and PEP-II.}]

\section{Introduction}

A scalar Higgs particle\cite{higgs} has been postulated over 30 years
ago as the mechanism of electroweak symmetry breaking in the Standard
Model (SM) of particle physics. This spontaneous breaking introduces
a huge hierarchy between the electroweak and Planck scales that is
unsatisfying. Extensions to the SM have been proposed over the years
to avoid unnatural fine-tuning. Supersymmetry\cite{susy} (SUSY) is one
such attractive extensions. Depending on its internal structure and
SUSY breaking mechanism, a variety of new phenomena are expected to be
observed. Rare signatures, as in high-mass tails or from SM suppressed
processes, are good places for generic beyond--the--Standard Model
searches.

The Large Electron Positron (LEP) collider at CERN completed operation
about four years ago. It ran at a center--of--mass energy of up to
$209 \, \mbox{GeV}$ and delivered about $1 \, \mbox{fb}^{-1}$ of data
to the four experiments, ALEPH, DELPHI, L3, and OPAL. The data are
analysed. Extensive searches for Higgs and new phenomena have come up
negative. For many new particles coupling to the Z boson LEP still
holds the most stringent limits.

Two machines, the Hadron Electron Ring Accelerator (HERA) and the
Tevatron, are currently running at the energie frontier with ever
increasing luminosities. HERA at DESY collides electrons or positrons
with protons at a center--of--mass energy of $319 \, \mbox{GeV}$. The
HERA upgrade increased the luminosity by a factor of 4.7. So far the
machine has delivered over $180 \, \mbox{pb}^{-1}$ of electron--proton
(about half) and positron--proton data to the two experiments, H1 and
ZEUS. The experiments are particularly sensitive to new particles
coupling to electron/positron and up/down quarks. HERA~II can also
deliver polarized lepton beams.

The Fermilab Tevatron collides proton and antiprotons at a center%
--of--mass energy of $1.96 \, \mbox{TeV}$. Luminosity
upgrades are continuing. So far the machine has delivered over
$1 \, \mbox{fb}^{-1}$ of data to the two experiments, CDF and D\O.
The improved detectors, higher cener--of--mass energy, and ten fold
increase in luminosity enable the experiments not only to
significantly extend previous searches but provide them with a
substantial discovery potential.

The Large Hadron Collider (LHC) at CERN and the International Linear
Collider (ILC) are machines under construction and in the planning
phase. The LHC will collide protons with protons at a center--of%
--mass energy of $14 \, \mbox{TeV}$. First collisions are expected
in 2007. The two experiments, ATLAS and CMS, have made detailed
studies of their reach to new physics. LHC is expected to boost our
sensitivity to new physics by an order of magnitude in energy/mass.
The ILC will collide electrons and positrons with a center--of--mass
energy of several hundred GeV. It will be the next generation machine
for precision measurments, like LEP was.

\subsection{Precision Electroweak and Top Measurments}

\begin{figure}
\epsfxsize68mm
\figurebox{}{}{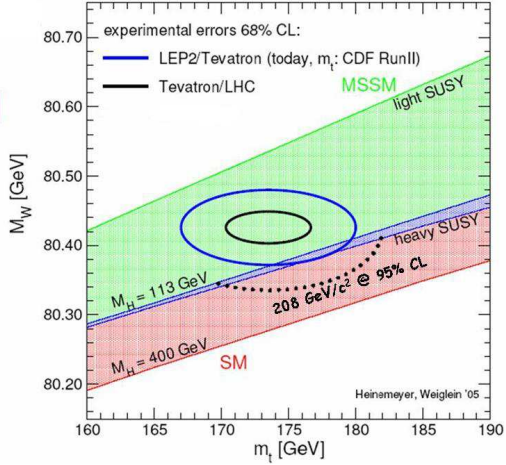}
\caption{One sigma contours of the current W and top mass measurments
compared to SM and MSSM Higgs masses.
Plot from Heinemeyer\protect\cite{heine} updated for new CDF top mass
measurment.}
\label{fig:mwmtmh}
\end{figure}

Precision electroweak measurments allow us to check the SM for
consistency or derive the mass of the unknown Higgs particle. For
Higgs prediction, the W boson mass and top quark mass are key
ingredients. With the new preliminary CDF Run~II top mass
measurment\cite{juste}, the world average is pulled down to
${\rm m}_{\rm t} = 174.3 \pm 3.4 \, \mbox{GeV}/c^2$.
Figure~\ref{fig:mwmtmh} shows the $1 \sigma$ and 95\% confidence
level (CL) contours of the W and top mass with overlaid Higgs mass.
Current measurments put the SM Higgs below $208 \, \mbox{GeV}/c^{2}$
at 95\% CL.

However, the top mass is an even more important ingredient for the
Higgs in Minimal Supersymmetric extensions of the Standard Model
(MSSM). The MSSM exclusion at low $\tan(\beta)$ derived from the
SM Higgs limit of LEP depends very sensitively on the mass of the
top quark and vanishes when the top mass is large.

\section{Standard Model Higgs}

The current lower limit on the Higgs mass of $114.4 \, \mbox{GeV}/c^{2}$
at 95\% CL comes from the LEP experiments\cite{lepsmh}. They did a fantastic
job of pushing the Higgs mass limit well above the Z pole where it would
be hard for proton--anti-proton experiments to detect. The Tevatron is the
current place for Higgs searches with an expected sensitivity to about
$130 \, \mbox{GeV}/c^{2}$. Here the main Higgs production mechanism is
via gluon-gluon fusion. Associated production with a W or Z has a factor
five lower cross-section. For low Higgs masses, below $135 \,
\mbox{GeV}/c^{2}$, the $\rm b \overline{b}$ decay mode is dominant. With
a leptonic W or Z decay we get signatures of zero, one, or two charged
leptons, an imbalance of energy in the transverse plane, missing
$E_{\rm T}$ (in case of zero or one charged lepton), and two b-jets.
For heavier Higgs the $\rm W W^{*}$ decay dominates and then Higgs
production via gluon fusion yields a viable signature.

\begin{figure}
\epsfxsize68mm
\figurebox{}{}{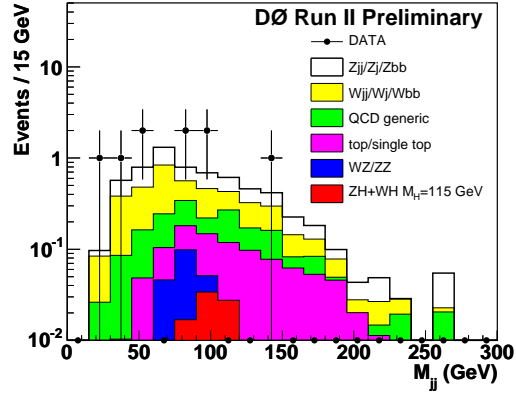}
\caption{Dijet mass spectrum of the D\O\ ZH analysis after two b-tags.}
\label{fig:d0zhmass}
\end{figure}

\begin{figure}
\epsfxsize68mm
\figurebox{}{}{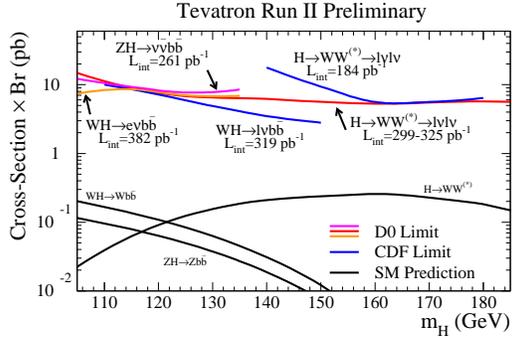}
\caption{Current CDF and D\O\ Run~II Higgs cross-section times branching
ratio limits from the WH, ZH, and $\rm W W^{*}$ channel.}
\label{fig:tevhiggs}
\end{figure}

The WH analyses of CDF and D\O\ were performed early on and results are
updated regularly with increased luminosity\cite{whana}. The D\O\ 
experiment has also completed a search in the ZH channel where the Z
decays into neutrinos\cite{zhana}. The analysis compares the missing
$E_{\rm T}$ (\mbox{$\not\!\!\!\!\;E_{\rm T}$})
as calculated from all energy in the detector with the calculation from
just clustered energy and the jet energy vector sum with the track
momentum vector sum to reduce instrumental background which comes mainly
from jet mismeasurments. The main background in the analysis comes from
Z plus multijet production and W plus $\rm b \overline{b}$ production
with W decay into $\tau \nu$ decay. Figure~\ref{fig:d0zhmass} shows the
dijet mass spectrum after two b-tags are required. No excess of events
over background expectation is observed in this search nor in any other
Higgs analysis of CDF and D\O. The cross-section times branching ratio
limit of this analysis is shown in Fig.~\ref{fig:tevhiggs} together with
the limits from the other Tevatron SM Higgs searches.

The sensitivity of CDF and D\O\ is currently between $3$ and $10 \,
\mbox{pb}$ while a SM Higgs is at about $0.2 \, \mbox{pb}$. The
difference between the current and the final Run~II Higgs sensitivity
projection\cite{tevh} is understood. In addition to the luminosity
accumulation, improvements in lepton and b-tagging acceptance, the
dijet mass resolution, and analysis techniques will bring the
sensitivity of the experiments to the projections made before Run~II.

\begin{figure}
\epsfxsize68mm
\figurebox{}{}{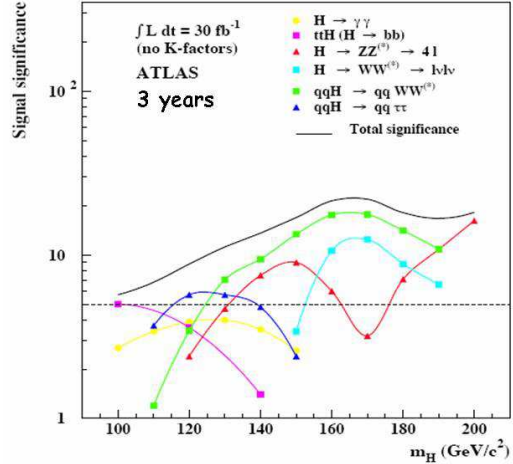}
\caption{SM Higgs signal significance of the ATLAS experiment for
the different search channels.}
\label{fig:lhchiggs}
\end{figure}

At LHC the Higgs production cross-section is huge. Even a decay mode
with small branching ratio, like Higgs into a photon pair, yields a
sizable event number. The two experiments each have an electromagnetic
calorimeter with very precise energy resolution to be able to observe
a diphoton mass bump from Higgs\cite{lhchgg} on top of the huge diphoton
continuum. For LHC vector boson fusion, however, will be the most
important production for Higgs. Both ATLAS and CMS can observe a SM
Higgs up to several hundred $\mbox{GeV}/c^{2}$ after a few years of
running, Fig.~\ref{fig:lhchiggs}. For LHC the observation of a Higgs
boson would be just the initial step. The two experiments can measure
the ratio of couplings and decay widths to an uncertainty of 20 and
30\%.

\section{MSSM Higgs}

Current and next generation experiments cover a SM Higgs well. The
Higgs sector, however, can be richer than a single doublet. Supersymmetry
extends the symmetry concept, that has been so successfull in particle
physics, to the spin sector. It provides a consistent framework for
gauge unification and solves the hierarchy problem of the SM. No SUSY
particles have been observed so far. Several SUSY breaking scenarios
are under consideration which determine the SUSY structure. The MSSM
is the general minimal supersymmetric extension of the SM. It has two
Higgs doublets yielding five physical Higgs particles: h, H, A,
$\rm H^{+}$, and $\rm H^{-}$. At tree level the Higgs sector is
described by two parameters, the pseudoscalar Higgs mass, $\rm m_{\rm A}$,
and ratio of the two Higgs vacuum expectation values, $\tan (\beta)$. The
MSSM, although the minimal extension, has a lot of free parameters. 
One normally uses models constrained based on SUSY breaking scenarios
and GUT scale relations or special benchmarking models.

At the Tevatron the Higgses of the MSSM are of particular interest.
The Yukawa coupling to down-type fermions is enhanced, boosting the
cross-section by a factor of $\tan(\beta)^{2}$. For large $\tan(\beta)$
the pseudoscalar Higgs and either h or H are expected to be almost
mass degenerate. The branching ratio into $\rm b \overline{b}$ is at
around 90\% independent of mass. Decays into tau pairs account for
close to 10\%.

\begin{figure}
\epsfxsize68mm
\figurebox{}{}{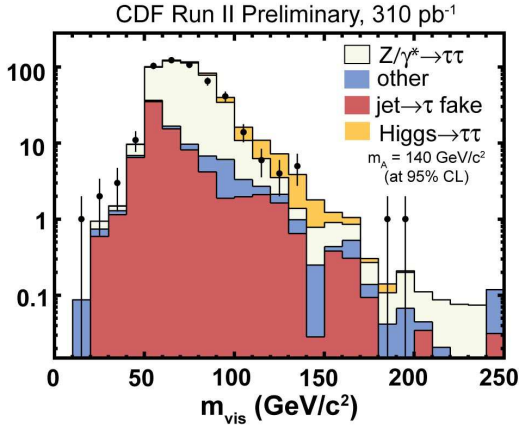}
\caption{Distribution of the ``visible'' Higgs mass in the CDF
ditau analysis.}
\label{fig:cdfhditaum}
\end{figure}

Two neutral MSSM Higgs searches are performed at CDF and D\O.
The first is based on Higgs plus $\rm b \overline{b}$ production:
$\rm b \overline{b} A \rightarrow b \overline{b} b \overline{b}$.
It yields a striking four b-jet signature. The second search is
based on the tau decay mode: $\rm A \rightarrow \tau^{+} \tau^{-}$.

\begin{figure}
\epsfxsize68mm
\figurebox{}{}{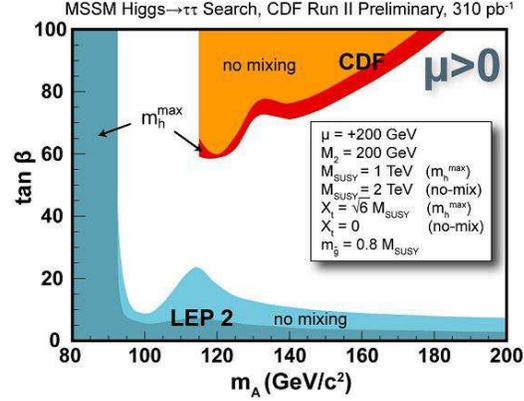}
\caption{Excluded regions in the mass of A versus $\tan(\beta)$
plane for the $\rm m_{h}^{\mbox{max}}$ and no mixing scenario
for the Higgsino mass marameter $\mu > 0$.}
\label{fig:cdfhtaupmu}
\end{figure}

Tau leptons are not as easily identified as electrons or muons. The CDF
analysis\cite{cdfhtau} is based on one leptonic tau decay and one
hadronic tau decay. Jets from hadronic tau decays are very narrow, pencil
like, compared to quark/gluon jets. CDF uses a double cone algorithm
to identify hadronic tau decays. An efficiency of 46\% is achieved
with a misidentification rate between 1.5\% to 0.1\% per jet depending
on the jet energy. For the Higgs search the experiment uses a data
sample selected by an electron or muon plus track trigger to achieve
high efficiency. Figure~\ref{fig:cdfhditaum} shows the visible mass
of the ditau system, calculated from the momentum vector of the lepton,
hadronic tau, and \mbox{$\not\!\!\!\!\;E_{\rm T}$}. The main background
comes from Z and Drell-Yan ditau production. No excess of events is observed
in the first $310 \, \mbox{pb}^{-1}$ of Run~II data. A binned likelihood
fit in the visible mass is used to set limits on the mass of A versus
$\tan(\beta)$, Fig~\ref{fig:cdfhtaupmu}.

\begin{figure}
\epsfxsize68mm
\figurebox{}{}{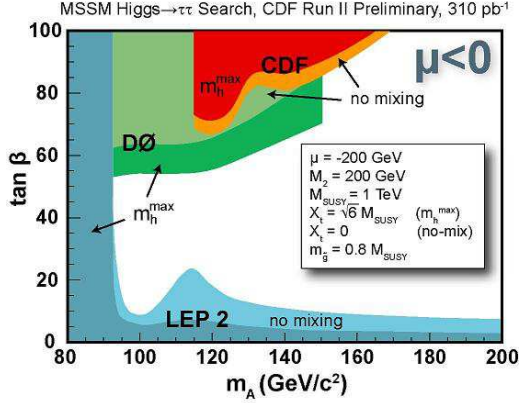}
\caption{Excluded regions in the mass of A versus $\tan(\beta)$
plane for the $\rm m_{h}^{\mbox{max}}$ and no mixing scenario
for the Higgsino mass marameter $\mu < 0$.}
\label{fig:cdfhtaunmu}
\end{figure}

The D\O\ analysis\cite{d0hbbb} for the four b-jet channel requires three
b-tag jets in the event. The first jet has to have $E_{\rm T} > 35 \,
\mbox{GeV}$ while the third can be as low as $15 \, \mbox{GeV}$.
To estimate the background from light quark and gluon jets, the
probability of mis-tagging a jet is measured on the three jet
sample before b-tagging, subtracting any true heavy flavour
contribution. Those mistag functions are then applied to the
untagged jets in the double b-tag sample to get the shape of the
multijet background to the triple b-tag sample. D\O\ determines
the overall background normalization by fitting the dijet mass
outside the hypothesized signal region in the triple b-tag sample.
Figure~\ref{fig:cdfhtaunmu} shows the mass versus $\tan(\beta)$
limit obtained by this analysis. For $\mu > 0$ the sensitivity
of the four b-jet channel is very low due to the lower cross-section
and lower branching ratio into $\rm b \overline{b}$, while for
the tau channel cross-section reduction and branching ratio enhancement
compensate.

\begin{figure}[tp]
\epsfxsize68mm
\figurebox{}{}{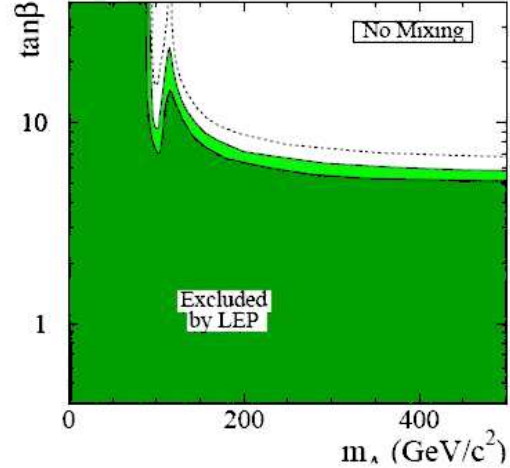}
\caption{Excluded regions in the mass of A versus $\tan(\beta)$
plane for the no mixing scenario. Dark areas are excluded at over
99\% CL, light areas at 95\% CL.}
\label{fig:lepmssmcl}
\end{figure}

\begin{figure}[bp]
\epsfxsize68mm
\figurebox{}{}{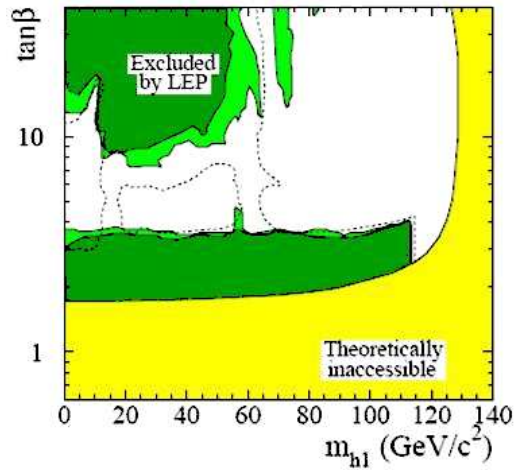}
\caption{Excluded regions in the mass of $\rm h_{1}$ versus $\tan(\beta)$
plane for the CP-violating scenario. Dark areas are excluded at over
99\% CL, light areas at 95\% CL.}
\label{fig:lepmssmcp}
\end{figure}

The final combined MSSM Higgs mass limits from the four LEP experiments
have been released\cite{lephmssm}. There are no signals of Higgsstrahlung
or pair production. Sensitivity is evaluated in several benchmark
models. A top mass of $179 \, \mbox{GeV}/c^{2}$ is assumed for all
limits. Figure~\ref{fig:lepmssmcl} shows the excluded mass of A versus
$\tan(\beta)$ for the classic no-stop mixing benchmark model with
${\rm M}_{\mbox{SUSY}} = 1000 \, \mbox{GeV}/c^{2}$, ${\rm M}_{2} =
200 \, \mbox{GeV}/c^{2}$, $\mu = -200 \, \mbox{GeV}/c^{2}$,
${\rm m}_{\mbox{gluino}} = 800 \, \mbox{GeV}/c^{2}$, $A = 0 + \mu \cdot
\cot(\beta)$. The
excluded area is reduced in the case of stop mixing and is quite
sensitive to the top mass. The LEP experiments also considered the
case of CP-violation in the Higgs sector. Such a scenario appeals
in explaining the cosmic matter--antimatter asymmetry. Experimentally
such a scenario is much more challenging as the lightest Higgs can
decouple from the Z. Figure~\ref{fig:lepmssmcp} shows the LEP results.
An inconsistency in the prediction from CPH and FeynHiggs for the
$\rm h_2 \rightarrow h_1 h_1$ branching ratio causes the hole at
$\tan(\beta) \sim 6$ to open up\cite{cphfhiggs}.

\begin{figure}
\epsfxsize68mm
\figurebox{}{}{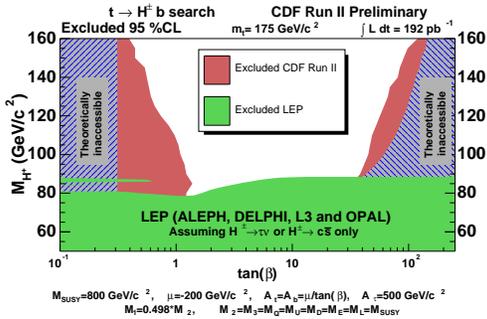}
\caption{Excluded regions in the $\tan(\beta)$ versus mass of charged
Higgs plane for a tevatron benchmarking scenario. The dark area is
excluded at 95\% CL, the dark line shows the expected limit with
$1 \sigma$ band.}
\label{fig:cdfhiggspm}
\end{figure}

There are also two charged Higgs particles in the MSSM. CDF uses
its top cross-section measurments from the various decay channels
to search for top decays into charged Higgs plus b-quark\cite{cdfhpm}.
Such a decay would change the expected number of events differently in
the dilepton, lepton plus single b-tag, lepton plus double b-tag, and
lepton plus tau channel, especially for small and large $\tan(\beta)$
values. Figure~\ref{fig:cdfhiggspm} shows the excluded mass as
function of $\tan(\beta)$ in one of the benchmark models studied.

\section{Supersymmetry}

From the LEP experiments\cite{lepchargino}we know that the chargino
has to be heavier than $103.5 \, \mbox{GeV}/c^{2}$. At the Tevatron
the cross-section for chargino--neutralino production is rather
small. However, in $R_{\rm P}$ conserving minimal supergravity
inspired SUSY (mSUGRA) one can get a very distinct signature.
In case of leptonic chargino and neutralino decay, the event
will contain only three charged leptons and missing $E_{\rm T}$
from the escaping neutrinos and the lightest SUSY particles (LSP).
The challenge in the analysis is the charged lepton acceptance
times efficiency since it enters with third power. For $\tan(
\beta)$ values above 8 to 10, tau decays become significant
and tau identification thus very important.

\begin{table}
\begin{center}
\caption{Observed events and expected number of background events
in the six channels of the D\O\ chargino--neutralino analysis.}
\label{tab:d0trilep}
\begin{tabular}{|l|c|c|} 
\hline 
 Channel                         & Expected        & Observed \\ \hline
 $\rm e \, e \, t$               & $0.21 \pm 0.12$ &  0 \\ \hline
 $\rm e \, \mu \, t$             & $0.31 \pm 0.13$ &  0 \\ \hline
 $\rm \mu \, \mu \, t$           & $1.75 \pm 0.57$ &  2 \\ \hline
 $\rm \mu^{\pm} \, \mu^{\pm}$    & $0.64 \pm 0.38$ &  1 \\ \hline
 $\rm e \, \tau_{h} \, t$        & $0.58 \pm 0.14$ &  0 \\ \hline
 $\rm \mu \, \tau_{h} \, t$      & $0.36 \pm 0.13$ &  1 \\ \hline \hline
 Total                           & $3.85 \pm 0.75$ &  4 \\
\hline
\end{tabular}
\end{center}
\end{table}

The D\O\ analysis\cite{d0chargino} searches in six separate channels
and combines the results. In all the channels known dilepton resonances
are removed and a combined cut on the \mbox{$\not\!\!\!\!\;E_{\rm T}$}
and $p_{\rm T}$ of the third lepton used to suppress background
from mainly misidentified leptons and diboson production.
Table~\ref{tab:d0trilep} shows the expected background and
observed number of events in each of the six channels. In the
$320 \, \mbox{pb}^{-1}$ of data analysed, no excess is observed.
D\O\ continues to set cross-section times branching ratio into
three lepton limits. The analysis also improves the LEP chargino
mass limit to $116 \, \mbox{GeV}/c^{2}$ in case of light sleptons,
i.e.\ small ${\rm m}_{0}$.

\begin{figure}
\epsfxsize68mm
\figurebox{}{}{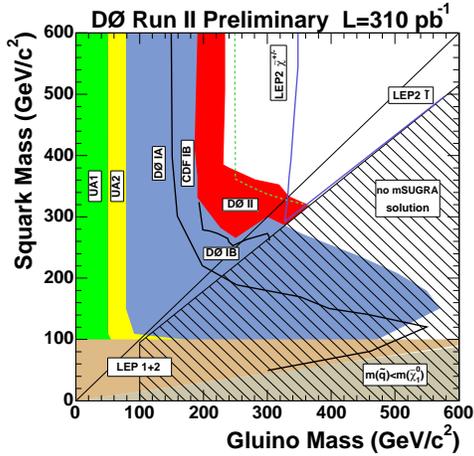}
\caption{Excluded region in the gluino versus squark mass plane of
the D\O\ missing $E_{\rm T}$ plus multijet analysis.}
\label{fig:d0jetmet}
\end{figure}

The production cross-section of coloured SUSY particles is much
larger than that of chargino--neutralino. The squarks of the first
two generations are assumed to be degenerate in mass. Stop and
sbottom quarks could be significantly lighter due to the large
top Yukawa coupling. CDF and D\O\ have dedicated analyses for
those\cite{tev3gen}. The analyses assume direct decay of the third
generation squark into LSP: $\tilde{\rm b} \rightarrow {\rm b}
\tilde{\chi}^{0}_{1}$ or $\tilde{\rm t} \rightarrow {\rm c}
\tilde{\chi}^{0}_{1}$. Both direct production of $\tilde{\rm t}
\overline{\tilde{\rm t}}$ and $\tilde{\rm b} \overline{\tilde{\rm b}}$
and indirect production through, for instance, $\tilde{\rm g}
\rightarrow \tilde{\rm b} \overline{\rm b}$ are researched.

For the gluinos and squarks of the first two generation the signature
depends on the mass hierarchy. If the squarks are lighter, squark
production is dominant and squarks will decay via $\rm \tilde{q}
\rightarrow q \tilde{\chi}^{0}$ or $\rm \tilde{q} \rightarrow q'
\tilde{\chi}^{\pm}$ with $\rm \tilde{\chi}^{\pm} \rightarrow q
\overline{q} \tilde{\chi}^{0}_{1}$. In case the gluino is the lighter
one, gluino pair production is dominant with $\rm \tilde{g}
\rightarrow q \overline{q} \tilde{\chi}^{0}$ or $\rm \tilde{g}
\rightarrow q \overline{q}' \tilde{\chi}^{\pm}$. With the jets from
the chargino decay generally being softer, this yields a 2, 3, or 4
jet signature together with missing $E_{\rm T}$. The
D\O\ analysis\cite{d0jetmet} makes a preselection, vetoing jet
back-to-back topologies, events with leptons, and events where the
missing $E_{\rm T}$ is close in azimuthal angle to a jet.
The main SM backgrounds are from W/Z plus multijet and QCD multijet
production. After the preselection dedicated analyses are made for
each jet multiplicity.  In all three cases the observed events are
explained by the background estimate. Figure~\ref{fig:d0jetmet} shows
the new excluded region in gluino versus squark mass.

\begin{figure}
\epsfxsize68mm
\figurebox{}{}{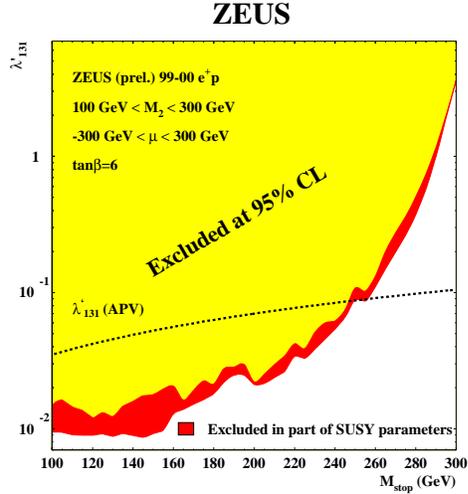}
\caption{Excluded stop mass as function of $\lambda'_{131}$ coupling
of the ZEUS analysis.}
\label{fig:zeusrpvsusy}
\end{figure}

\begin{figure}
\epsfxsize68mm
\figurebox{}{}{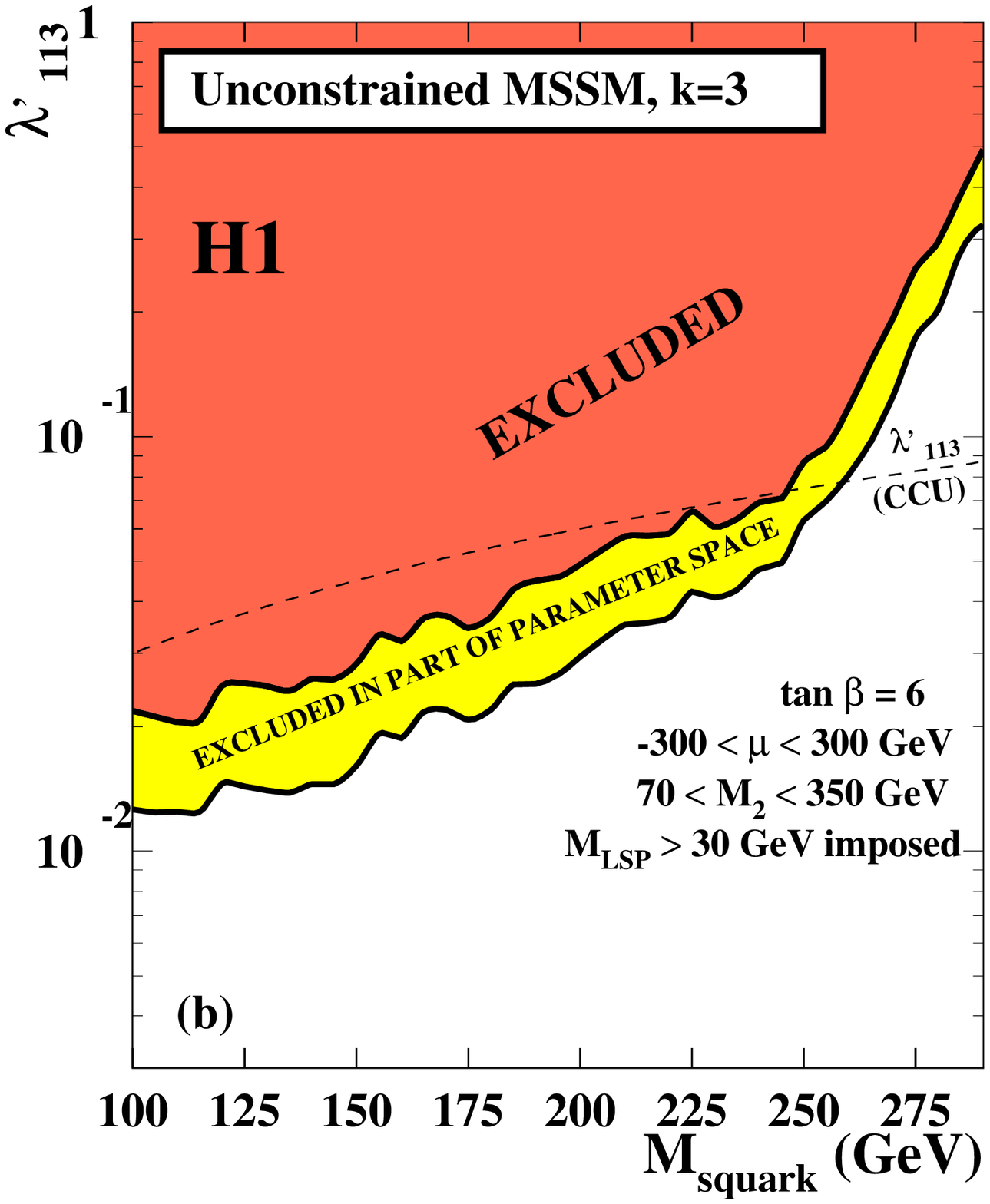}
\caption{Excluded sbottom mass as function of $\lambda'_{113}$ coupling
of the H1 analysis.}
\label{fig:h1rpvsusy}
\end{figure}

$R_{\rm P}$ conserving SUSY yields a natural dark matter candidate.
However, $R$-parity conservation is really put into the models {\it ad
hoc} and nature may not conserve it. In case of $R_{\rm P}$ violation
(\mbox{$\not\!\!\!\,R_{\rm P}$}), different signatures arise. In the case
of a non-vanishing $\lambda'$ coupling electrons and u/d-quarks can
couple, ideal for HERA. In electron--proton mode, $\lambda'_{11k}$
couplings are accessible while in positron--proton mode, $\lambda'_{1j1}$
couplings would produce resonant $\rm \tilde{u}_{L}$. Both H1 and ZEUS
searched for a large variety of \mbox{$\not\!\!\!\,R_{\rm P}$} signatures.
The most striking signature is "wrong" sign electrons, i.e. events with
energetic electrons while in positron--proton mode and events with
positrons, jets, and no missing $E_{\rm T}$ while in electron--proton
mode. No signals of \mbox{$\not\!\!\!\,R_{\rm P}$} SUSY has been found.
Figures~\ref{fig:zeusrpvsusy} and \ref{fig:h1rpvsusy} show the stop
and sbottom mass limits as function of the coupling constant from ZEUS
and H1. Squarks with $R_{\rm P}$ violating couplings of electroweak
strength are excluded up to $275 \, \mbox{GeV}/c^{2}$.

In case SUSY is broken via gauge interactions (GMSB) the gravition
acquires a small mass and becomes the LSP. The next-to-lightest SUSY
particle will decay into a photon plus gravitino for the distinct
GMSB photon signature. CDF and D\O\ have both searched in the diphoton
plus \mbox{$\not\!\!\!\!\;E_{\rm T}$} channel\cite{tevgmsb}. The experiments
use chargino--neutralino production as reference model for the search.
The two experiment have combined their results from the first $250 \,
\mbox{pb}^{-1}$ of data and exclude charginos in GMSB models below
$209 \, \mbox{GeV}/c^{2}$ at 95\% CL.

\begin{figure}
\epsfxsize68mm
\figurebox{}{}{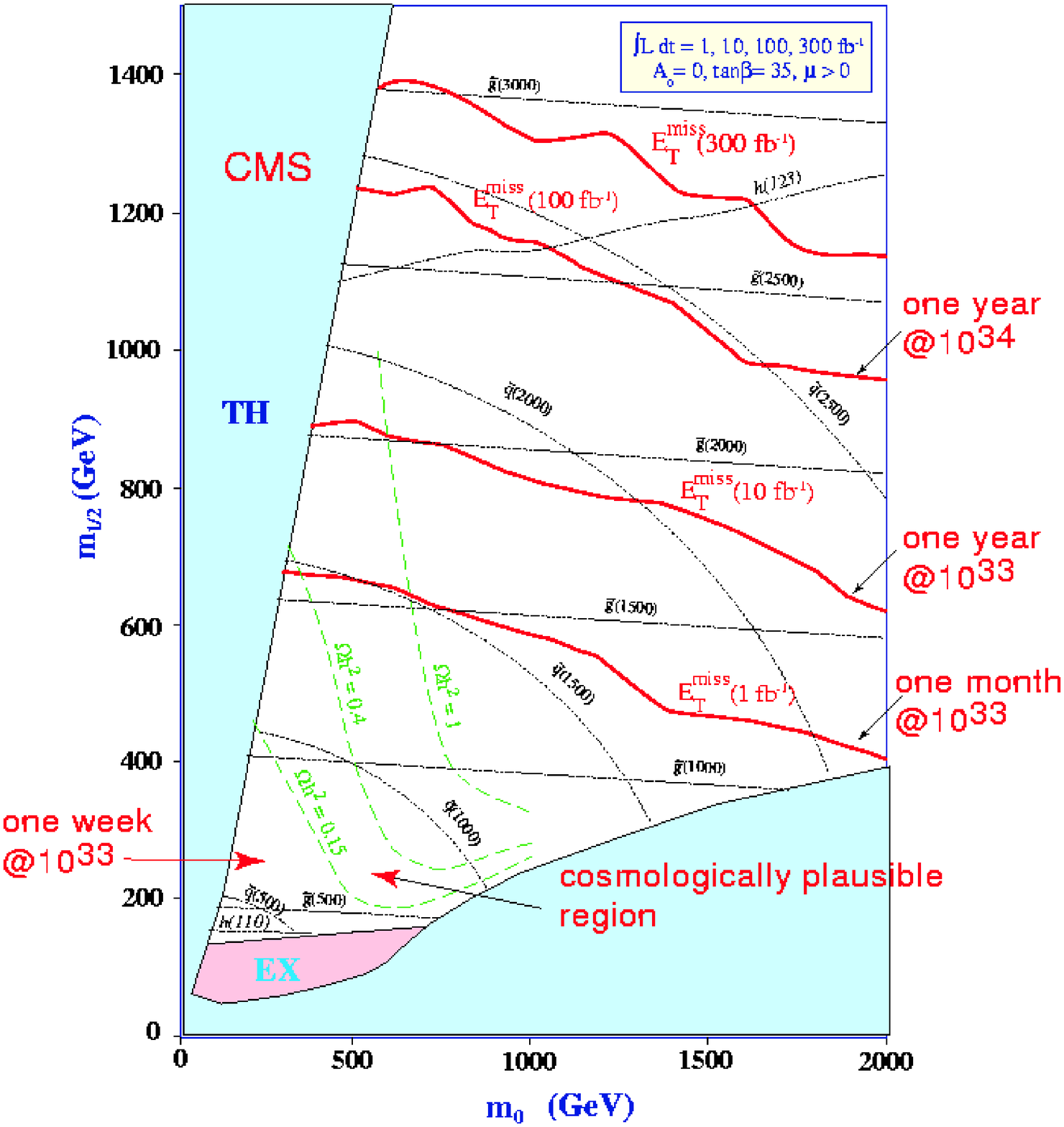}
\caption{SUSY sensitivity of CMS in the $\rm m_{0}$ versus $\rm m_{1/2}$
plane for different integrated luminosity.}
\label{fig:lhcsusy}
\end{figure}

In the case of low energy supersymmetry, LHC will be a great
machine\cite{lhcsusy}. It will provide a definite answer to the question
and with a small luminosity of only a few month probe SUSY scales of over
a TeV, Figure~\ref{fig:lhcsusy}. But LHC can do more and measure sparticle
masses, for instance, for the second lightest neutralino from the
dilepton spectrum endpoint or even the gluino mass from the
top--bottom endpoint.

The ILC\cite{ilcsusy}, however, will be required for precision mass
and coupling measurements.

\section{Isolated Lepton and Missing Energy}

In Run~I of HERA H1 observed an excess of events with isolated lepton
$p_{\rm T} > 10 \, \mbox{GeV}$ and missing $E_{\rm T} > 12 \, \mbox{GeV}$
beyond what one would expect from W production\cite{herah1ex}. The excess
was pronounced at large $p_{\rm T}^{X} > 25 \, \mbox{GeV}$ and did not
fit well any new physics model. Both H1 and ZEUS have searched for an
isolated lepton plus \mbox{$\not\!\!\!\!\;E_{\rm T}$} signature in the new
Run~II data\cite{heraisol}. H1 has used the identical selection in
the analysis of the new data, separately for positron--proton and
electron--proton data. The muon channel shows no more the excess seen in
Run~I while an event excess remains in the electron channel. H1 has also
analysed the tau data from Run~I which show no excess either. ZEUS did
not observe an event excess in Run~I. With $40 \, \mbox{pb}^{-1}$ of
Run~II data analysed ZEUS finds also no excess in the electron channel.
Table~\ref{tab:heraisol} shows the current results of all the searches.

\begin{table}
\begin{center}
\epsfxsize68mm
\figurebox{}{}{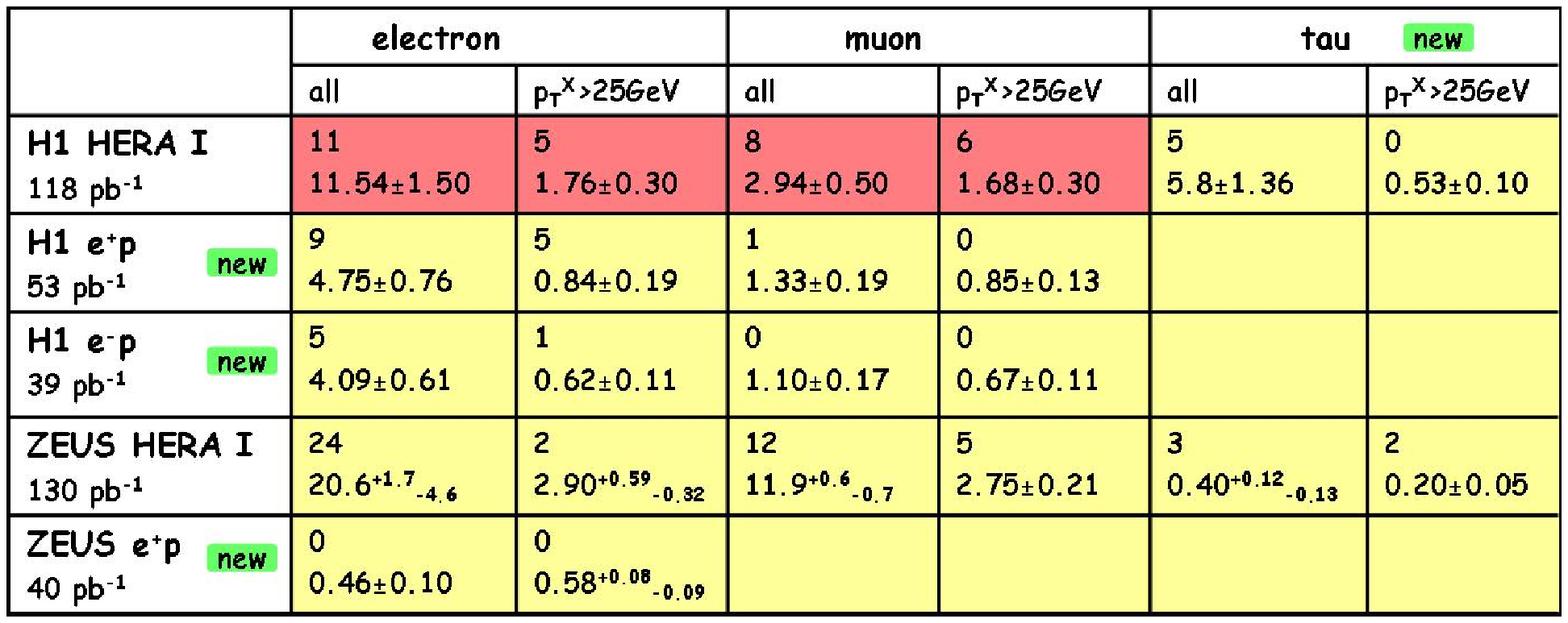}
\caption{Expected and observed number of isolated electron, muon, and
tau events of H1 and ZEUS.}
\label{tab:heraisol}
\end{center}
\end{table}

\section{High Mass Searches}

High-mass searches were one of the first results presented from Run~II
of the Tevatron. New gauge bosons and other high mass resonances yield
energetic objects when they decay. Searches based on energetic leptons,
photons, and missing $E_{\rm T}$ give access to a large variety of new
physics. For instance, events with an energetic electron and positron
are sensitive to Z', large extra dimensions, Randall-Sundrum gravitons,
\mbox{$\not\!\!\!\!\;R_{\rm P}$} sneutrinos, and technicolor particles,
$\rho$ and $\omega$. The analyses\cite{tevhighm} of CDF and D\O\ are
constantly refined, on one side to cover signatures in a generic way, by
for instance calculating sensitivity based on the spin, or to incorporate
new models and interpretations, like expressing Z' sensitivity based
on d-{\it x}u, or B-{\it x}L couplings, on the other side to include
additional event kinematics like $\cos \theta^{*}$ in the analysis to enhance
sensitivity to new physics. About $450 \, \mbox{pb}^{-1}$ of Run~II
data are analysed for high mass objects. No excess or deviation are
observed so far.

\section{Indirect Searches}

With no signals of new physics in any of the direct searches, we
can search for signs of new physics where new particles are in virtual
states. Processes that are rare in the SM provide an excellent place to
search for signs of new physics.

\begin{figure}
\epsfxsize68mm
\figurebox{}{}{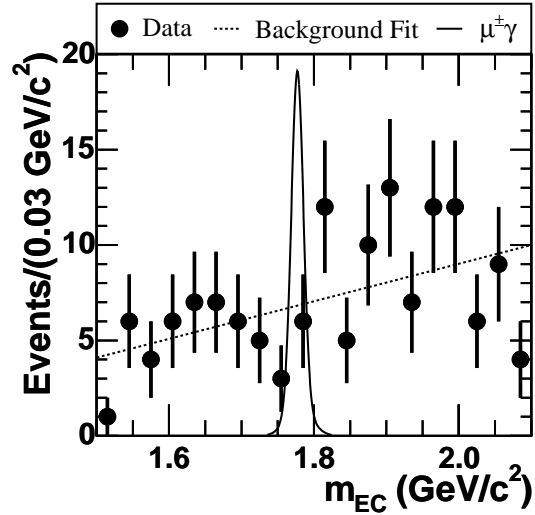}
\caption{The energy constraint muon photon mass spectrum of the BaBar
tau decay into muon plus photon analysis.}
\label{fig:babarmec}
\end{figure}

Tau decays into a muon and a photon are tiny in the SM with a branching
ratio around $10^{-40}$ but allowed if one includes neutrino mixing.
The decay violates lepton flavour which occures naturally in SUSY grand
unified theories. Both Belle and BaBar\cite{lfvtau} have recorded over
20 million ditau events. BaBar uses one tau as tag and then the
other as probe. A neural network is used to discriminate signal from
background. The main background comes from dimuon production and
ditau production with tau decays into a muon plus neutrinos and a photon
from initial or final state radiation. Figure~\ref{fig:babarmec} shows
the energy constraint muon photon mass with a curve of how a potential
signal would look. Observation agrees with the background expectation
and BaBar sets a 90\% CL limit on the branching ratio of tau into a muon
plus a photon at $6.8 * 10^{-8}$.

\begin{figure}
\epsfxsize68mm
\figurebox{}{}{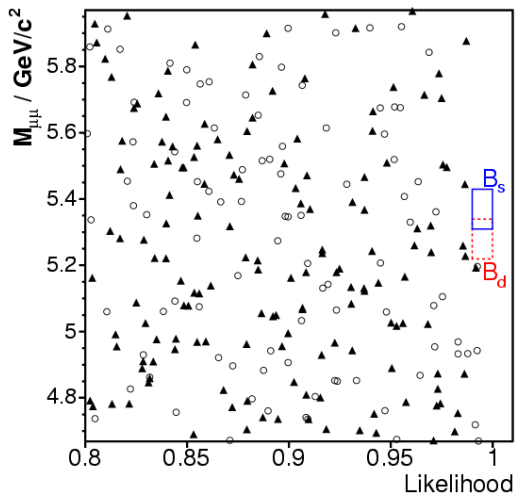}
\caption{Distribution of events in the likelihood versus dimuon mass
plane of the CDF $\rm B \rightarrow \mu^{+} \mu^{-}$ analysis.}
\label{fig:cdfbmumu}
\end{figure}

Another interesting channel is the $\rm B_{s}$ into $\mu^{+} \mu^{-}$
decay. The flavor changing neutral current (FCNC) decay is heavily
suppressed in the SM. In the MSSM, however, the branching ratio is
enhanced, proportional to $\tan(\beta)^{6}$. CDF has a long tradition
of searching for ${\rm B} \rightarrow \mu^{+} \mu^{-}$. The analysis is
normalized to the observed $\rm B^{+} \rightarrow J/\psi K^{+}$ decays
to become independent of the b production cross-section. A likelihood
function is used to separate dimuons that originate from a decay of a
particle with lifetime from prompt dimuons. CDF observes no events in
the $\rm B_{d}$ and $\rm B_{s}$ window, Fig.~\ref{fig:cdfbmumu}. The
combined CDF/D\O\ analyses set a 95\% CL branching ratio limit of
$1.2 * 10^{-7}$ for $\rm B_{s}$ and $3.1 * 10^{-8}$ for $\rm B_{d}$.
This excludes first regions in SUSY parameter space at high $\tan(\beta)$.

\section{Summary and Outlook}

Scientists have explored nature to smaller and smaller scales over the
years. In the last 50 years particle physics has made tremendous progress,
revealing and exploring the next smaller layer of particles. We have
developed a self-consistent, although incomplete, model that describes
our current knowledge. Nature still surprises us, like with the observation
of neutrino oscillation and the accelerating expansion of the universe.
Our current understanding strongly suggests new physics to be close to the
electroweak scale. However, no significant evidence of new physics has
been observed so far. The current experiments search extensively in a
large variety of signatures for deviations from the Standard Model. Some
of the most interesting and promising search channels were presented in
this review. Both HERA and the Tevatron are running well with record
luminosities and the experiments are keeping up analysing the data. The
hope is on the current experiments to unveil the next layer or the next
symmetry of nature. A new generation of experiments is only a few years
away and should answer our question about new electroweak scale physics.
The transfer of expertise and experience to those new experiments has
started.

\end{document}